\documentclass[aip,jcp,twocolumn,reprint]{revtex4-1}

\usepackage{graphics,graphicx,amsfonts,amsmath,amsbsy,amssymb,color}
\usepackage{bm}
\usepackage[a4paper,vmargin={20mm,20mm},hmargin={20mm,10mm}]{geometry}
\usepackage{subfigure}
\usepackage{paralist}
\usepackage{bbold}
\usepackage{mathtools, nccmath}

\newcommand{\bra}{\langle}
\newcommand{\ket}{\rangle}

\def \refWF {{| \phi_m^{(0)} \ket}}
\def \refWFn {{ \phi_m^{(0)} }}
\def \refE {{E_m^{(0)}}}
\def \Slk {{S_l^{(k)}}}
\def \Sref {{S_0^{(0)}}}
\def \Elk {{E_l^{(k)}}}
\def \ElkHat {{\hat{E}_l^{(k)}}}
\def \Nlk {{N_l^{(k)}}}
\def \psilk {{\psi_l^{(k)}}}
\def \ELDn {{E_L^{\mathrm{D}}[n]}}
\def \ELDmean {{\overline{E_L^{\mathrm{D}}}}}
\def \Nnorm {{N_{\mathrm{norm}}}}
\def \Ninit {{N_{\mathrm{init}}}}
\def \Nenergy {{N_{\mathrm{energy}}}}
\def \NElk {{N_{\mathrm{\Elk}}}}
\def \ncarbon {{N}}
\def \quintet {{ {}^5 \mathrm{A}_{\mathrm{g}} }}
\def \triplet {{ {}^3 \mathrm{B}_{1\mathrm{g}} }}
\def \Cu2O2 {{ [\mathrm{Cu}_2\mathrm{O}_2]^{2+} }}

\begin{document}

\title{Efficient multireference perturbation theory without high-order reduced density matrices}
\author{Nick S. Blunt}
\email{nicksblunt@gmail.com}
\affiliation{Department of Chemistry, Lensfield Road, Cambridge, CB2 1EW, United Kingdom}
\author{Ankit Mahajan}
\affiliation{Department of Chemistry, University of Colorado, Boulder, CO 80302, USA}
\author{Sandeep Sharma}
\email{sanshar@gmail.com}
\affiliation{Department of Chemistry, University of Colorado, Boulder, CO 80302, USA}

\begin{abstract}
We present a stochastic approach to perform strongly contracted $n$-electron valence state perturbation theory (SC-NEVPT), which only requires one- and two-body reduced density matrices, without introducing approximations. We use this method to perform SC-NEVPT2 for CASSCF wave functions obtained from selected configuration interaction, although the approach is applicable to a larger class of wave functions, including those from variational Monte Carlo (VMC). The accuracy of this approach is demonstrated for small test systems, and the scaling is investigated with the number of virtual orbitals and the molecule size. We also find the SC-NEVPT2 energy to be relatively insensitive to the quality of the reference wave function. Finally, the method is applied to the Fe(II)-Porphyrin system with a $(32\mathrm{e}, 29\mathrm{o})$ active space, and to the isomerization of $ \Cu2O2 $ in a $(28\mathrm{e}, 32\mathrm{o})$ active space.
\end{abstract}

\date{\today}

\maketitle

\section{Introduction}
\label{sec:intro}

In studying electronic structure problems, correlation effects are often separated into strong and dynamic correlation. In some systems, a single determinant is sufficient to provide a qualitative description of a system's electronic structure. However, in strongly correlated system this assumption breaks down, and one often has to use a superposition of multiple determinants to describe the reference state. These determinants are obtained by including all (or several) possible occupations within a subset of orbitals known as the active space. One often optimizes the active space orbitals to minimize the energy which results in a method known as complete active space self consistent field (CASSCF). The rest of the correlation due to excitation into remaining orbitals is known as dynamic correlation and can be included using a variety of methods including multireference configuration interaction (MRCI)\cite{Werner1988, Knowles1992}, multireference perturbation theory (MRPT)\cite{Andersson1990, Angeli2001, Angeli2002} and multireference coupled cluster (MRCC)\cite{Bartlett2007, Evangelista2018}.

In recent years there have been significant improvements in algorithms for performing (near-exact) CASSCF calculations. Methods including the density matrix renormalization group algorithm (DMRG)\cite{White1992, Chan2002}, full configuration interaction quantum Monte Carlo (FCIQMC)\cite{Booth2009, Cleland2010, Thomas2015_3, Manni2016} and selected configuration interaction (SCI)\cite{Huron1973, Evangelisti1983, Garniron2017, Smith2017}, can now be used to solve CASSCF problems accurately for active spaces of $40$ to $50$ orbitals, and possibly beyond\cite{Dominika2008, Ghosh2008, Thomas2015_3, Manni2016, Smith2017}. However, there still remains the important task of including dynamic correlation. Traditional implementations of MRCI and MRPT requires calculating and storing the three and sometimes four-body reduced density matrices (RDMs) within the active space, which require $\mathcal{O}(n_a^6)$ and $\mathcal{O}(n_a^8)$ storage, respectively. This becomes infeasible for the large active spaces considered above, and separate approaches must be developed.

A variety of methods have been proposed and used to avoid the need for higher-order RDMs. These include the use of cumulant approximations\cite{Zgid2009}; uncontracting terms that require high-order RDMs\cite{Celani2000}; use of matrix product states\cite{Sharma2014_2, Sharma2017_2}; approaches based on FCIQMC (where the high-order RDMs are only sampled)\cite{Anderson2020, Halson2020}; and others\cite{gagliardi, pernal, giner, Chenyang2019, Deustua2018}.

Recently, we demonstrated that it is possible to perform strongly contracted MRCI (SC-MRCI) and second-order $n$-electron valence perturbation theory (SC-NEVPT2) without constructing RDMs, but instead using variational Monte Carlo (VMC)\cite{Mahajan2019}. In this approach, rather than constructing RDMs and contracting them with integrals, it is possible to directly sample contributions from determinants in the first-order interacting space (FOIS). Although the number of determinants in the FOIS grows exponentially with the active space size,  VMC provides a polynomial scaling method to sample them. These stochastic approaches were referred to as SC-MRCI(s) and SC-NEVPT2(s).

In this article, we develop this idea further, focussing specifically on SC-NEVPT2(s). In particular, we present a somewhat different algorithm that is more efficient, and avoids the need for a trial wave function in the FOIS, as was required in the original SC-NEVPT2(s) approach. We also extend this to include core orbitals, which were not considered in our original implementation. We go on to provide analysis of this approach, including scaling with the number virtual orbitals for the N$_2$ molecule, and with system size for polyacetylene molecules. We also provide examples demonstrating the performance of SC-NEVPT2 when the reference wave function is in error. We then study two much larger systems than considered previously by this method, namely Fe(II)-Porphyrin in a $(32\mathrm{e}, 29\mathrm{o})$ active space, and $ \Cu2O2 $ in a $(28\mathrm{e}, 32\mathrm{o})$ active space, demonstrating that this approach is practical for challenging problems.

\section{SC-NEVPT2 overview}
\label{sec:nevpt_overview}

We begin by recapping the strongly-contracted NEVPT2 method\cite{Angeli2001, Angeli2002, Angeli2001_2}, and defining notation to be used throughout.

In multireference perturbation theory, one begins by solving the complete active space (CAS) problem, giving a reference wave function $\refWF$,
\begin{equation}
\refWF = \sum_I C_{I,m} | D_I \ket,
\end{equation}
where $m$ is the state label and $|D_I\rangle$ are determinants in which all core orbitals (denoted $i$, $j$, $\ldots$) are occupied, all virtual orbitals (denoted $r$, $s$, $\ldots$) are unoccupied, while active orbitals (denoted $a$, $b$, $\ldots$) can take any occupation number.

Assuming that the set of active orbitals is chosen appropriately, $\refWF$ provides a qualitative description of the true wave function. For better accuracy, dynamic correlation must then be included by considering excitations involving core and virtual orbitals. This can be done by second-order perturbation theory, after choosing an appropriate reference Hamiltonian ($\hat{H}_0$). There is no unique way of defining a reference Hamiltonian, in fact, any Hamiltonian that has $|\phi_m^{(0)}\rangle$ as its ground state can be selected. Various reference Hamiltonians have been chosen in literature and each leads to a different perturbation theory. In this article we take the reference Hamiltonian to be the one which defines strongly-contracted NEVPT (SC-NEVPT) theory (see Equation~\ref{eq:ham}).

In the SC scheme, the uncontracted FOIS is partitioned into subspaces $\Slk$. Here, $k$ specifies the change in the number of active space electrons relative to $\refWF$ ($-2 \le k \le 2$), while $l$ specifies which non-active orbitals are involved in the excitation. For example, a determinant which contains two unoccupied core orbitals $i$ and $j$, and a single occupied virtual orbital $r$ belongs to class $S_{ij,r}^{(1)}$. A single perturber state is then assigned to each class $\Slk$, defined by
\begin{equation}
| \psi_{l}^{(k)} \ket = P_l^{(k)} H \refWF,
\end{equation}
where $P_l^{(k)}$ is the projector onto the $S_l^{(k)}$ subspace and $H$ is the Hamiltonian operator. This definition ensures that perturber states are orthogonal to each other (but not normalized).

The above perturber states can be further divided into eight types, depending on the number of core and virtual orbitals involved. We refer to these as: $v$, $vv$, $c$, $cv$, $cvv$, $cc$, $ccv$ and $ccvv$. For example, we say that a perturber state $| \psi_{ij,r}^{(1)} \ket$ is of type $ccv$.

Given the perturber states $| \psi_{l}^{(k)} \ket$, the zeroth-order Hamiltonian for SC-NEVPT is defined as 
\begin{equation}
H^{(0)} = \sum_m E_m^{(0)} | \phi_m^{(0)} \ket \bra \phi_m^{(0)} | + \sum_{l,k} E_l^{(k)} | \psi_{l}^{(k)} \ket \bra \psi_{l}^{(k)} |\label{eq:ham},
\end{equation}
which leads to the second-order perturbative energy correction,
\begin{equation}
E_m^{(2)} = \sum_{l,k} \frac{ N_l^{(k)} }{E_m^{(0)} - E_l^{(k)}}.
\label{eq:nevpt2_energy}
\end{equation}
Here, $N_l^{(k)}$ are the squared norms of the perturbers,
\begin{equation}
N_l^{(k)} = \bra \psi_{l}^{(k)} | \psi_{l}^{(k)} \ket.
\end{equation}
$E_m^{(0)}$ is the zeroth-order energy for state $m$, and $E_l^{(k)}$ are the perturber energies. In NEVPT these perturber energies are defined via the Dyall Hamiltonian, $H^D$,
\begin{equation}
E_l^{(k)} = \frac{1}{N_l^{(k)}}\bra \psi_{l}^{(k)} | H^D | \psi_{l}^{(k)} \ket,
\end{equation}
with
\begin{equation}
H^D = \sum_i^{\mathrm{core}} \epsilon_i a_i^{\dagger} a_i + \sum_a^{\mathrm{virtual}} \epsilon_a a_a^{\dagger} a_a + H_{\mathrm{active}},
\end{equation}
where $H_{\mathrm{active}}$ is the core-averaged Hamiltonian in the active space, such that $H^{D} \refWF = E_m^{(0)} \refWF$.

The primary task is to calculate the second-order energy from Eq.~(\ref{eq:nevpt2_energy}). To do so, both the squared norms, $N_l^{(k)}$, and the perturber energies, $E_l^{(k)}$, are required. Exact expressions for $E_l^{(k)}$ and $N_l^{(k)}$ can be obtained in terms of active-space RDMs. However, these include three and four-body RDMs, whose storage requirements scale as $\mathcal{O}(n_a^6)$ and $\mathcal{O}(n_a^8)$ in the number of active-space orbitals, $n_a$. Instead, we will take a stochastic approach which avoids the need for higher-order RDMs.

\section{Stochastic SC-NEVPT2}
\label{sec:stoch_nevpt}

Estimation of $E_m^{(2)}$ can be performed in two stages. First we calculate the squared norms, $\Nlk$ for all perturbers. In the second step we calculate the summation in Equation~\ref{eq:nevpt2_energy} stochastically by sampling perturbers $| \psilk \ket$ with probabilities proportional to $\Nlk$. For the selected perturber we estimate the energy $\Elk$ and accumulate the contribution towards $E_m^{(2)}$ (this is fully described in Section~\ref{sec:stoch_energy}).

We begin with some important general points. First, we only wish to avoid the use of three and four-body RDMs; storing one and two-body RDMs is always straightforward for current active spaces. We therefore use the existing SC-NEVPT2 approach to calculate all instances of $\Elk$ and $\Nlk$ which only require the 1-RDM and 2-RDM. This greatly reduces the sampling task to be performed. Using this rule, in the stochastic approach to be described, we can ignore $ccvv$, $cvv$ and $ccv$ contributions entirely. For $cc$, $cv$ and $vv$ we need to sample $\Elk$, but not $\Nlk$. For the remaining two sets ($c$ and $v$), both $\Elk$ and $\Nlk$ must be sampled. This is summarised in Table~\ref{tab:exact_stoch}.

For the algorithm to be presented, the only computational requirement on $\refWF$ is that overlaps such as $\bra n | \refWFn \ket$ can be calculated, and that the 1-RDM and 2-RDM can be constructed. In this article we solely take $\refWF$ from SCI. However, this requirement is met by other wave functions, such as matrix product states, or those used in VMC.

Whenever generating connections by application of the Hamiltonian, it should be understood that the heat bath criteria is applied. This was described in the initial presentation of our VMC approach, which we refer to for in-depth description\cite{Sabzevari2018}. This ensures that, for a determinant $|n\ket$, connections $|p\ket$ are not generated if $ | \bra p | \hat{H} | n \ket | < \epsilon $, for some small threshold $\epsilon$. For results in this article we always take $\epsilon = 10^{-8}$ Ha. This typically reduces both the prefactor and scaling of the resulting algorithm, with a negligible effect on the accuracy.

\begin{table}[t]
{\footnotesize
\begin{tabular}{@{\extracolsep{4pt}}lcc@{}}
\hline
\hline
Perturber type & Energies ($\Elk$) & Norms ($\Nlk$) \\
\hline
$c$     &  Stochastic   &  Stochastic  \\
$v$     &  Stochastic   &  Stochastic  \\
$cc$    &  Stochastic   &  Exact       \\ 
$cv$    &  Stochastic   &  Exact       \\
$vv$    &  Stochastic   &  Exact       \\
$ccv$   &  Exact        &  Exact       \\
$cvv$   &  Exact        &  Exact       \\
$ccvv$  &  Exact        &  Exact       \\
\hline
\hline
\end{tabular}
}
\caption{Table showing which $\Elk$ and $\Nlk$ instances are calculated stochastically, and which are calculated by the traditional deterministic approach. The deterministic approach is taken if only $1$ and $2$-body RDMs are required, otherwise we use the stochastic approach in order to avoid $3$ and $4$-body RDMs.}
\label{tab:exact_stoch}
\end{table}

\subsection{The Continuous Time Monte Carlo algorithm}
\label{sec:ctmc}

In the following, it is necessary to sample from probability distributions $\rho_n$, which take the form
\begin{equation}
\rho_n = \frac{ | \bra n | \psi \ket |^2 }{ \bra \psi | \psi \ket },
\end{equation}
where $|\psi\ket$ is some wave function. Typically in VMC, this would be sampled by the Metropolis-Hastings algorithm\cite{Metropolis1953, Ceperley1977, Foulkes2001}. However, this can be quite inefficient when working in a discrete basis of Slater determinants. Instead we use the Continuous Time Monte Carlo (CTMC) algorithm\cite{Bortz1975, Gillespie1976}, which was introduced to VMC recently\cite{Sabzevari2018}. When applied in other areas, this algorithm is sometimes known as Kinetic Monte Carlo (KMC) or the Bortz-Kalos-Lebowitz (BKL) algorithm. We briefly recap it here:
\begin{enumerate}
\item From a determinant $| n \ket$, calculate $r(p \leftarrow n)$,
\begin{equation}
r(p \leftarrow n) = \bigg| \frac{ \bra p | \psi \ket }{ \bra n | \psi \ket } \bigg|,\label{eq:ovlpratio}
\end{equation}
for all determinants $| p \ket$ connected to $| n \ket$ by a single or double excitation (within the relevant space).

\item Calculate the residence time for $| n \ket$, defined as
\begin{equation}
t_n = \frac{1}{\sum_p r(p \leftarrow n)}.
\end{equation}
This will define the weight of contributions from $|n\ket$ in subsequent estimators.

\item Select a new determinant $|p\ket$ with probability proportional to $r(p \leftarrow n)$.
\end{enumerate}
After a short burn-in period, iterating this procedure will correctly sample $\rho_n$, provided that $t_n$ are used as weights for contributions to estimators. We denote the total residence time for a random walk by $T = \sum_n t_n$. It is worth pointing out that in CTMC all moves are accepted and there are no rejections, but this comes at the added cost of having to evaluate all the overlap ratios in Equation~(\ref{eq:ovlpratio}). However, this additional cost is mitigated in our VMC algorithm because these overall ratios are obtained when evaluating the local energy.

\subsection{Sampling $N_l^{(k)}$}
\label{sec:stoch_norm}

We take the general case where $\refWF$ may not be normalized. The squared norms can be sampled using the following approach:
\begin{align}
N_l^{(k)} &= \frac{ \bra \psi_l^{(k)} | \psi_l^{(k)} \ket }{ \bra \refWFn | \refWFn \ket }, \\
          &= \frac{ \bra \refWFn | \hat{H} \hat{P}_l^{(k)} \hat{H} | \refWFn \ket }{ \bra \refWFn | \refWFn \ket }, \\
          &= \sum_{n \in S_0^{(0)}} \frac{ | \bra n | \refWFn \ket |^2 }{ \bra \refWFn | \refWFn \ket } \frac{ \bra n | \hat{H} \hat{P}_l^{(k)} \hat{H} | \refWFn \ket }{ \bra n | \refWFn \ket }, \\
          &= \Big\langle N_l^{(k)}[n] \Big \rangle_{\rho_n}.
\end{align}
Here, $\rho_n$ is the probability distribution to be sampled by a random walk,
\begin{equation}
\rho_n = \frac{ | \bra n | \refWFn \ket |^2 }{ \bra \refWFn | \refWFn \ket }.
\end{equation}
The determinants selected, $| n \ket$, are referred to as walkers. We emphasize that this random walk takes places entirely within the CASCI space ($S_0^{(0)}$). The quantity $N_l^{(k)}[n]$ is defined by
\begin{align}
N_l^{(k)}[n] &= \frac{ \bra n | \hat{H} \hat{P}_l^{(k)} \hat{H} | \refWFn \ket }{ \bra n | \refWFn \ket }, \\
             &= \frac{ \sum_{p \in S_l^{(k)}} \bra n | \hat{H} | p \ket \sum_{r \in S_0^{(0)}} \bra p | \hat{H} | r \ket \bra r | \refWFn \ket }{ \bra n | \refWFn \ket }.
\label{eq:nlk_n}
\end{align}
$N_l^{(k)}[n]$ is calculated by the following steps. First, generate all determinants $| p \ket$ in $S_l^{(k)}$ that are connected to $| n \ket$ (calculating $\bra n | \hat{H} | p \ket$ for each). Then for each $| p \ket$, generate all connected determinants $ | r \ket$ within $S_0^{(0)}$ (calculating $\bra p | \hat{H} | r \ket$ and $\bra r | \refWFn \ket$ for each).

In practice, instead of calculating $N_l^{(k)}[n]$ for each $S_l^{(k)}$ separately, we accumulate all instances simultaneously. That is, for each walker $| n \ket \in S_0^{(0)}$, loop over connected determinants $| p \ket$ in all $S_l^{(k)}$ being considered, accumulating contributions to $N_l^{(k)}[n]$ for each.

The norm and energy of the zeroth-order wave function are sampled in an analogous way during the same random walk. In this article we take $\refWF$ from SCI, such that the wave function is normalized by construction, and its energy known. Nonetheless, this step is important in general.

Walker moves within $S_0^{(0)}$ are made using the continuous time Monte Carlo (CTMC) algorithm, described above. Importantly, each $r(p \leftarrow n)$ is already constructed in order to obtain $\bra \refWFn | \refWFn \ket$, so that the CTMC algorithm can be performed essentially for free.

Note that for every $S_l^{(k)}$ sampled, the quantity $\bra p | \psi_l^{(k)} \ket = \bra p | \hat{H} | \refWFn \ket$ is calculated for at least one determinant $| p \ket$ in $S_l^{(k)}$. We can therefore keep a list of determinants which have the largest value of $\bra p | \psi_l^{(k)} \ket$ for each $S_l^{(k)}$ sector (of the determinants reached). These determinants are used to initialize the walkers when sampling the corresponding $\Elk$.

As noted in Table~\ref{tab:exact_stoch}, we only need to sample norms for perturbers of type $c$ and $v$. However, we also need to generate initial determinants for $cc$, $cv$ and $vv$. Therefore, there are two parameters which specify the sampling in this step, which we denote $\Nnorm$ and $\Ninit$. For the first $\Nnorm$ iterations, $\Nlk$ is only sampled for $c$ and $v$-type perturbers. We then perform $\Ninit$ iterations in which $\Nlk$ is sampled for all $5$ perturber types ($c$, $v$, $cc$, $cv$ and $vv$). The $\Nlk$ estimates for $cc$, $cv$ and $vv$ from this step are \emph{not} used, as we have access to the exact values. Instead, we use the generated initial determinants when sampling $E_l^{(k)}$, in the next step. These final iterations are more expensive. However, we always take $\Ninit \ll \Nnorm$, and typically $\Ninit = 50$ is more than sufficient.

\subsection{Sampling $E^{(2)}$ and $E_l^{(k)}$}
\label{sec:stoch_energy}

We next consider the sampling of $E_m^{(2)}$ itself, as defined in Eq.~(\ref{eq:nevpt2_energy}). This is done by sampling terms in this summation with a probability proportional to $\Nlk$:
\begin{align}
E^{(2)} &= \sum_{k,l \ne 0} \frac{1}{E^{(0)} - E_l^{(k)}} N_l^{(k)}, \\
        &= \Big[ \sum_{k',l' \ne 0} N_{l'}^{(k')} \Big] \times \sum_{k,l \ne 0} \frac{1}{E^{(0)} - E_l^{(k)}} \cdot \frac{ N_l^{(k)} }{ \sum_{k',l' \ne 0} N_{l'}^{(k')} }, \\
        &= \Big[ \sum_{k,l \ne 0} N_l^{(k)} \Big] \times \Big \langle \frac{1}{E^{(0)} - E_l^{(k)}} \Big \rangle_{\rho(l,k)},
\label{eq:energy_sampling}
\end{align}
where $\rho(l,k) = \frac{ N_l^{(k)} }{ \sum_{k',l' \ne 0} N_{l'}^{(k')} }$. It is straightforward to sample from $\rho(l,k)$, since all $N_l^{(k)}$ values are stored after the initial stage of the algorithm. We also truncate the summation to only include contributions with $N_l^{(k)} \ge 10^{-8}$, as an efficiency improvement which we do not find to affect the accuracy.

For each $S_l^{(k)}$ selected, the corresponding $E_l^{(k)}$ must then be estimated. This is achieved by a random walk entirely within $S_l^{(k)}$. Specifically,
\begin{align}
E_l^{(k)} &= \frac{ \bra \psi_l^{(k)} | \hat{H}_{\mathrm{D}} | \psi_l^{(k)} \ket }{ \bra \psi_l^{(k)} | \psi_l^{(k)} \ket }, \\
          &= \sum_{n \in \Slk} \frac{ | \bra n | \psi_l^{(k)} \ket |^2 }{ \bra \psi_l^{(k)} | \psi_l^{(k)} \ket } \frac{ \bra n | \hat{H}_{\mathrm{D}} | \psi_l^{(k)} \ket }{ \bra n | \psi_l^{(k)} \ket }, \\
          &= \Big\langle E_L^{\mathrm{D}}[n] \Big\rangle_{\rho_n},
\end{align}
where
\begin{equation}
\rho_n = \frac{ | \bra n | \psi_l^{(k)} \ket |^2 }{ \bra \psi_l^{(k)} | \psi_l^{(k)} \ket }
\end{equation}
and $E_L^{\mathrm{D}}[n]$ is the local energy at $| n \ket$ with respect to $ \hat{H}_{\mathrm{D}} $:
\begin{align}
E_L^{\mathrm{D}}[n] &= \frac{ \bra n | \hat{H}_{\mathrm{D}} | \psi_l^{(k)} \ket }{ \bra n | \psi_l^{(k)} \ket }, \label{eq:local_dyall} \\
                    &= \frac{ \sum_{p \in S_l^{(k)}} \bra n | \hat{H}_{\mathrm{D}} | p \ket \sum_{r \in S_0^{(0)}} \bra p | \hat{H} | r \ket \bra r | \refWFn \ket }{ \sum_{r \in S_0^{(0)}} \bra n | \hat{H} | r \ket \bra r | \refWFn \ket }.
\label{eq:elk_n}
\end{align}
The numerator of this expression is calculated by the following steps. First, generate all connections $| p \ket$ within $S_l^{(k)}$ (and calculate each $\bra n | \hat{H}_{\mathrm{D}} | p \ket$). Then for each $ | p \ket $, generate all connections $| r \ket$ within $S_0^{(0)}$ (and calculate each $\bra p | \hat{H} | r \ket$ and $ \bra r | \refWFn \ket$). Similarly, the denominator of this expression is obtained by looping over all connected determinants $|r\ket$ in $S_0^{(0)}$, and calculating $\bra n | \hat{H} | r \ket$ and $ \bra r | \refWFn \ket$ for each.

The distribution $\rho_n$ is again sampled using the CTMC algorithm. All required values of $r(p \leftarrow n)$ are obtained when $E_L^{\mathrm{D}}[n]$ is calculated, such that this can be performed essentially for free.

There are two parameters which define the sampling in this step, which we denote $\Nenergy$ and $\NElk$. Here, $\Nenergy$ is the number of samples taken from $\rho(l,k)$ (i.e., the number of $\Elk$ selected), while $\NElk$ is the number of samples to estimate each $\Elk$ selected. However, instead of using a fixed iteration count for all $\Elk$, it is often more accurate to use a fixed residence time instead (see Appendix~\ref{sec:biases}). In cases where we use a fixed residence time, we will list both the total residence time used, denoted $T$, and also the average iteration count per $\Elk$ estimate.

\subsection{Parallelism}

The above algorithm can be efficiently performed on large-scale computers. In our current implementation, this is done by running the above steps independently on each MPI process. Each process generates its own $\Nlk$ and $\Elk$ estimates, and ultimately its own $E^{(2)}$ estimate at the end of the simulation. These $E^{(2)}$ values are then averaged to produce the final estimate of the SC-NEVPT2 energy, together with an error estimate. This error estimate is simple to obtain, since results from different processes are statistically independent. There is no communication between MPI processes at any point during the simulation.

This approach has very good parallel efficiency. The only cause of non-ideal parallel performance is that processes will take varying times to complete all iterations.

Note that the sampling parameters defined above ($\Nnorm$, $\Ninit$, $\Nenergy$ and $\NElk$) are the number of iterations performed on \emph{each} process.

\begin{table}[t]
{\footnotesize
\begin{tabular}{@{\extracolsep{4pt}}lcc@{}}
\hline
\hline
Perturber type & Energies ($\Elk$) & Norms ($\Nlk$) \\
\hline
$c$     &  $\mathcal{O}(n_a^7)$    &  $\mathcal{O}(n_a^6 n_c)$      \\
$v$     &  $\mathcal{O}(n_a^7)$    &  $\mathcal{O}(n_a^6 n_v)$      \\
$cc$    &  $\mathcal{O}(n_a^6)$    &  $\mathcal{O}(n_a^4 n_c^2)$    \\ 
$cv$    &  $\mathcal{O}(n_a^6)$    &  $\mathcal{O}(n_a^4 n_c n_v)$  \\
$vv$    &  $\mathcal{O}(n_a^6)$    &  $\mathcal{O}(n_a^4 n_v^2)$    \\
\hline
\hline
\end{tabular}
}
\caption{The expected scaling to sample $\Elk$ or $\Nlk$ estimates. The scaling of $\Elk$ is for a fixed $(l,k)$, while for $\Nlk$ is for all $(l,k)$ of a given type. This assumes that all valid excitations are generated, whereas excitations are actually generated by the heat bath criteria, which is expected to reduce scaling. However, the number of samples required to maintain a constant statistical error will usually increase with system size, increasing the overall scaling. Scaling for real examples is investigated in Section~\ref{sec:results}.}
\label{tab:scaling_theor}
\end{table}

\subsection{Scaling}

In the following, we denote the number of core, active and virtual orbitals as $n_c$, $n_a$ and $n_v$, respectively.

In the algorithm presented, a norm estimate $\Nlk$ is obtained for all $(l,k)$ for which the heat bath criteria is satisfied. However, only a subset of $\Elk$ are obtained, as sampled according to the distribution in Eq.~(\ref{eq:energy_sampling}). We therefore consider the scaling to calculate $\Nlk$ for \emph{all} $(l,k)$ values, and to calculate $\Elk$ for a constant number of $(l,k)$ samples.

Consider the cost to calculate all $\Nlk[n]$, for a given $| n \ket \in \Sref$. The expression to be evaluated is given in Eq.~(\ref{eq:nlk_n}). First, all determinants $ | p \ket \notin \Sref $ connected to $| n \ket$ are generated. For perturbers of type $c$, $v$, $cc$, $cv$ and $vv$, the number of valid $ | p \ket$ scales as $\mathcal{O}(n_a^3 n_c)$, $\mathcal{O}(n_a^3 n_v)$, $\mathcal{O}(n_a^2 n_c^2)$, $\mathcal{O}(n_a^2 n_c n_v)$ and $\mathcal{O}(n_a^2 n_v^2)$, respectively. For each $ | p \ket \in \Slk$, the cost to generate all connected $ | r \ket \in S_0^{(0)}$ then scales as $\mathcal{O}(n_a^3)$ for $c$ and $v$-type perturbers, and as $\mathcal{O}(n_a^2)$ for $cc$, $cv$ and $vv$-type perturbers.

$\Elk[n]$ is calculated by Eq.~(\ref{eq:elk_n}). For a given $ | n \ket \in \Slk$, the cost to generate all connected $ | p \ket \in \Slk $ scales as $\mathcal{O}(n_a^4)$ for all perturber types. Then, for each $ | p \ket \in \Slk $, the cost to generate all connected $ | r \ket \in S_0^{(0)} $ scales as $\mathcal{O}(n_a^3)$ for $c$ and $v$-type perturbers, and as $\mathcal{O}(n_a^2)$ for $cc$, $cv$ and $vv$-type perturbers.

The overall scaling for each perturber type, obtained from the above arguments, is given in Table~\ref{tab:scaling_theor}. The true scaling will be somewhat different to this in practice. First, we do not loop over all connected determinants, but instead use the heat bath criteria, where connections are not generated if they have a Hamiltonian element below some threshold. This is expected to reduce the overall scaling (however, in this article we use CASSCF orbitals; because these are delocalized, the potential benefits are more limited). Second, the above only gives the scaling to calculate $\Elk[n]$ and $\Nlk[n]$ for a constant number of samples, $| n \ket$. In general, the number of samples will increase with system size, for a constant statistical error. This increases the overall scaling. It is difficult to write down a general formula to describe this effect. We instead investigate this through examples in Section~\ref{sec:results}.

\section{Results}
\label{sec:results}

In the following, PySCF\cite{pyscf, pyscf2020} is used to generate molecular orbitals via CASSCF, and to generate molecular integrals for the subsequent SC-NEVPT2(s) calculations. Heat bath CI (HCI) as implemented in the Dice code is used as the CASSCF solver\cite{Holmes2016_2, Sharma2017, Smith2017}, and also to generate the zeroth-order wave function $\refWF$ for the SC-NEVPT2(s) step.

To account for burn-in errors, we discard the initial $50$ iterations for each CTMC random walk, both for $\Nlk$ and $\Elk$ estimation, unless stated otherwise.

\subsection{Scaling with number of virtual orbitals: N$_2$}
\label{sec:scaling_vir}

\begin{table*}[t]
\begin{center}
{\footnotesize
\begin{tabular}{@{\extracolsep{4pt}}rccccccc@{}}
\hline
\hline
 & & & & & & \multicolumn{2}{c}{Total energy $+ 109$ (Ha)} \\
\cline{7-8}
Basis & $n_v$ & $t_{\mathrm{norm}}$ (s) & $t_{\mathrm{init. \: det.}}$ (s) & $t_{\mathrm{energy}}$ (s) & Statistical error (Ha) & SC-NEVPT2(s) & Molpro SC-NEVPT2 \\
\hline
cc-pVDZ     &   18  &   5.103  &   1.019  &  234.110  &  $2.0 \times 10^{-4}$  &  -0.1857(2)  &  -0.18543  \\ 
aug-cc-pVDZ &   36  &  10.482  &   2.454  &  238.149  &  $2.2 \times 10^{-4}$  &  -0.2025(2)  &  -0.20236  \\
cc-pVTZ     &   50  &  15.672  &   3.989  &  204.581  &  $3.5 \times 10^{-4}$  &  -0.2844(4)  &  -0.28498  \\
aug-cc-pVTZ &   82  &  26.841  &   8.526  &  225.526  &  $3.6 \times 10^{-4}$  &  -0.2946(4)  &  -0.29433  \\
cc-pVQZ     &  100  &  30.979  &  10.756  &  221.810  &  $5.2 \times 10^{-4}$  &  -0.3452(5)  &  -0.34511  \\
aug-cc-pVQZ &  150  &  52.517  &  21.578  &  231.923  &  $5.0 \times 10^{-4}$  &  -0.3494(5)  &  -0.34880  \\
cc-pV5Z     &  172  &  56.710  &  25.333  &  224.181  &  $4.1 \times 10^{-4}$  &  -0.3680(4)  &  -0.36752  \\
aug-cc-pV5Z &  244  &  94.875  &  50.289  &  257.278  &  $3.0 \times 10^{-4}$  &  -0.3706(3)  &  -0.36956  \\
cc-pV6Z     &  270  &  99.372  &  55.366  &  244.260  &  $4.9 \times 10^{-4}$  &  -0.3828(5)  &  -0.38285  \\
aug-cc-pV6Z &  368  & 152.935  & 101.737  &  271.484  &  $4.6 \times 10^{-4}$  &  -0.3838(5)  &  -0.38406  \\
\hline
\hline
\end{tabular}
}
\caption{Scaling of SC-NEVPT2(s) timing and error estimates with basis set size, applied to the ground state of N$_2$ at $R = 2.5$ $\mathrm{a_0}$. The active space is $(10\mathrm{e}, 8\mathrm{o})$. $t_{\mathrm{norm}}$ is the time to perform $900$ iterations to sample $\Nlk$ for $c$ and $v$-type perturbers. $t_{\mathrm{init. \: det.}}$ is the time to perform $100$ iterations to generate initial determinants. $t_{\mathrm{energy}}$ is the time to sample $10,000$ values of $\Elk$. The final two columns compare the subsequent SC-NEVPT2(s) energy estimates to exact results from Molpro\cite{MOLPRO}.}
\label{tab:n2}
\end{center}
\end{table*}

As a simple first example, we consider N$_2$ in its ground state at $R = 2.5$ $\mathrm{a_0}$ bond length. The active space is $(10\mathrm{e}, 8\mathrm{o})$, with $2$ core orbitals. We then consider calculating the SC-NEVPT2 energy for increasing correlation consistent basis sets, from cc-pVDZ ($18$ virtual orbitals) to aug-cc-pV6Z ($368$ virtual orbitals).

For the norm-sampling stage, we use parameters $\Nnorm = 900$ and $\Ninit = 100$. For the energy sampling stage, we use $\Nenergy = 10,000$ and $\NElk = 100$.

Results are presented in Table~\ref{tab:n2}. The final two columns compare the stochastic SC-NEVPT2 energies to those calculated with Molpro\cite{MOLPRO}, which agree within $1$ or $2$ statistical error bars.

Timing and error results from Table~\ref{tab:n2} can be used to assess scaling with respect to the number of virtual orbitals. Based on the theoretical scaling in Table~\ref{tab:scaling_theor}, and for a \emph{fixed number of iterations}, one would except the sampling of norms (time $t_{\mathrm{norm}}$) to asymptotically scale with the number of virtual orbitals as $\mathcal{O}(n_v)$. The expected asymptotic scaling to generate initial determinants (time $t_{\mathrm{init. \: det.}}$) is $\mathcal{O}(n_v^2)$. Sampling a constant number of energies (time $t_{\mathrm{energy}}$) should be independent of $n_v$.

\begin{figure}[t]
\includegraphics[width=\linewidth]{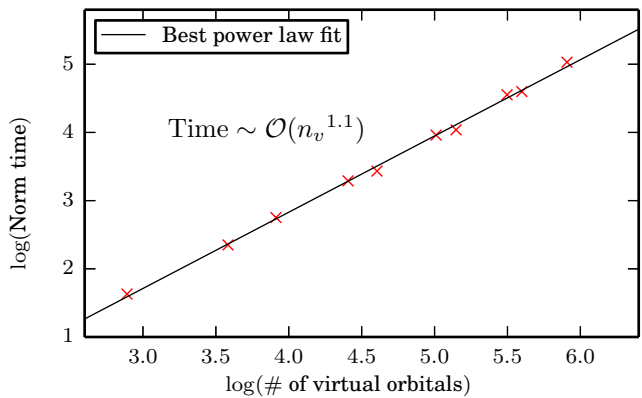}
\caption{Scaling of the norm-sampling time ($t_{\mathrm{norm}}$) against the number of virtual orbitals. The system is N$_2$ at $R = 2.5$ $\mathrm{a_0}$, with a $(10\mathrm{e}, 8\mathrm{o})$ active space. A constant number of iterations are performed, $\Nnorm = 900$. The scaling is found to be $t_{\mathrm{norm}} \sim \mathcal{O}(n_v^{1.1})$.}
\label{fig:norm_fit}
\end{figure}

$t_{\mathrm{energy}}$ is seen to be independent of $n_v$ as expected. Meanwhile, the observed scaling of $t_{\mathrm{norm}}$ is $\mathcal{O}(n_v^{1.1})$, while the observed scaling of $t_{\mathrm{init. \: det.}}$ is $\mathcal{O}(n_v^{1.5})$, in reasonable agreement with the predicted results. The scaling and fit for $t_{\mathrm{norm}}$ is shown in Figure~\ref{fig:norm_fit}.

It is more challenging to reason about how quickly the statistical error should increase. There are two sources of statistical error, first from sampling the norms, and second from sampling the energies. We usually observe that the majority of statistical error comes from the energy sampling step, though this will depend on how the simulation parameters are chosen. In the present case, there is a noticeable increase in statistical error from cc-pVDZ to cc-pVQZ, but interestingly the error becomes somewhat insensitive to $n_v$ beyond this point. There is an error on each of these error estimates, but these are small enough to not affect this conclusion.

\subsection{Scaling with molecule length: Polyacetylene}
\label{sec:scaling_len}

\begin{table*}[t]
\begin{center}
{\footnotesize
\begin{tabular}{@{\extracolsep{4pt}}lcccccccc@{}}
\hline
\hline
$\#$ of C atoms ($N$)  & $n_c$  & $n_a$  & $n_v$  & $n_{\mathrm{dets}}$  & $t_{\mathrm{norm}}$ (s) & $t_{\mathrm{init. \: det.}}$ (s) & $t_{\mathrm{energy}}$ (s) & Statistical error (Ha) \\
\hline
4   & 9   & 4   & 31   & 20        &  1.556                  &  1.165   &  3.469      & $1.1 \times 10^{-4} $  \\
8   & 17  & 8   & 59   & 2458      &  26.469                 &  29.072  &  90.156     & $2.2 \times 10^{-4} $  \\
12  & 25  & 12  & 87   & $7.9 \times 10^4$   &  208.94                 &  251.76  &  865.784    & $3.0 \times 10^{-4} $  \\
16  & 33  & 16  & 115  & $4.2 \times 10^5$   &  $1.134 \times 10^{3}$  &  $1.279 \times 10^{3}$ & $5.086 \times 10^{3}$  &  $4.8 \times 10^{-4} $  \\
20  & 41  & 20  & 143  & $1.4 \times 10^6$   &  $4.694 \times 10^{3}$  &  $4.832 \times 10^{3}$ & $2.006 \times 10^{4}$  & $6.8 \times 10^{-4} $  \\ 
24  & 49  & 24  & 171  & $2.5 \times 10^6$   &  $1.981 \times 10^{4}$  &  $1.594 \times 10^{4}$ & $6.733 \times 10^{4}$  & $4.7 \times 10^{-4} $  \\
28  & 57  & 28  & 199  & $3.3 \times 10^6$   &  $5.615 \times 10^{4}$  &  $4.222 \times 10^{4}$ & $1.822 \times 10^{5}$  & $1.1 \times 10^{-3} $  \\
\hline
\hline
\end{tabular}
}
\caption{Simulation time and statistical error for SC-NEVPT2(s) simulations performed on polyacetylene, as the number of carbon atoms ($N$) is increased. $n_c$, $n_a$ and $n_v$ give the number of core, active and virtual orbitals, respectively. A constant number of iterations was performed for each simulation (see main text for simulation parameters).}
\label{tab:poly}
\end{center}
\end{table*}

To consider scaling with overall molecule size, we consider trans-polyacetylene molecules with two terminal hydrogen atoms. These take the form C$_{2n}$H$_{2n+2}$. We denote the number of carbon atoms as $N$, and consider cases from $\ncarbon=4$ to $\ncarbon=28$. The corresponding number of core, active and virtual orbitals are given in Table~\ref{tab:poly}.

The orbital basis set is 6-31g. This is not large enough for accurate quantitative results, but sufficient for the present scaling study. Similarly, we take a model geometry, where all bond lengths and angles are fixed. Specifically, single C-C bond lengths are $1.45$ \AA, double C-C bond lengths are $1.34$ \AA, and C-H bond lengths are $1.08$ \AA. All angles are set to $120^\circ$.

For larger values of $\ncarbon$, the CASCI problem becomes infeasible to solve by FCI. Instead we use selected CI (SCI), specifically the heat bath CI (HCI) method. A constant HCI threshold of $\epsilon = 5 \times 10^{-5}$ Ha is used for each value of $\ncarbon$. The number of determinants in the HCI wave function is reported as $n_{\mathrm{dets}}$ in Table~\ref{tab:poly}.

The same parameters are used for each simulation: $\Nnorm = 900$, $\Ninit = 100$, $\Nenergy = 1000$ and $\NElk = 100$. Simulations were run on 32 cores on two Intel E5-2650 nodes.

Each of $n_c$, $n_a$ and $n_v$ scale linearly with the number of carbon atoms. Therefore, from Table~\ref{tab:scaling_theor}, the idealised asymptotic scaling for a constant number of iterations is $\mathcal{O}(\ncarbon^7)$. If we discard the $\ncarbon=4$ data point (to better investigate the asymptotic scaling), then the observed scaling for the total time time ($t_{\mathrm{norm}} + t_{\mathrm{init. \: det.}} + t_{\mathrm{energy}}$) is $\mathcal{O}(\ncarbon^{6.1})$. This lower scaling is reasonable, given that the theoretical scaling does not account for excitations ignored by the heat bath criteria.

There is also an increase in the final statistical error with molecule size. Interestingly, this error decreases from $\ncarbon=20$ to $\ncarbon=24$; we have checked that this is accurate, and not the result of error on the error estimate. However, all other data points follow the expected trend of increasing error.

\begin{figure}[t]
\includegraphics[width=\linewidth]{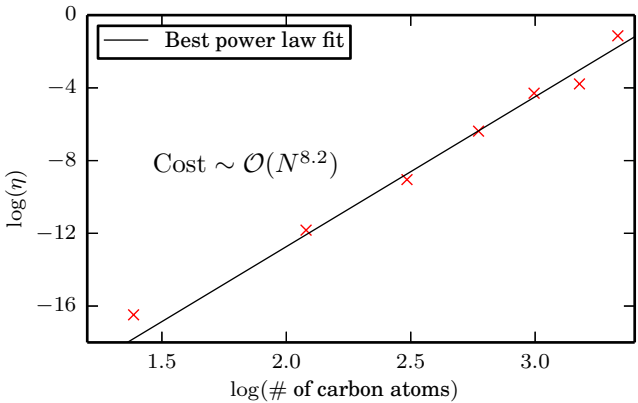}
\caption{A measure of computational cost in SC-NEVPT2(s), plotted against the number of carbon atoms ($N$) in polyacetylene molecules. The cost is $\eta = \sigma \times t^2$, where $\sigma$ is the statistical error, and $t$ the total simulation time. The cost is seen to scale roughly as $\mathcal{O}(N^{8.2})$.}
\label{fig:poly_fit}
\end{figure}

Statistical error decreases with the number of samples ($n_s$) as $n_s^{-1/2}$, and so decreases with simulation time ($t$) as $t^{-1/2}$. Therefore, a sensible measure of overall computational cost is
\begin{equation}
\eta = t \times \sigma^2,
\end{equation}
where $t$ is the total time and $\sigma$ is the final error estimate. For	the polyacetylene data in Table~\ref{tab:poly}, the values of $\eta$ are plotted in Figure~\ref{fig:poly_fit}, which agree well with a linear regression line on this log-log plot. Excluding the first data point ($\ncarbon=4$), the overall cost scales roughly as $\mathcal{O}(\ncarbon^{8.2})$. Although this scaling is steep, it is similar to that of traditional SC-NEVPT2, but with the benefit of not requiring higher-order RDMs. In the next sections, we demonstrate that the method is feasible for active spaces with $32$ orbitals. Given the favorable parallel efficiency, we expect active spaces with more than $40$ orbitals to be achievable. Nonetheless, we are investigating alternative approaches to reduce this scaling.

\subsection{Effect of error in the reference wave function}
\label{sec:ref_error}

It is interesting to investigate how the accuracy of the reference wave function affects the final SC-NEVPT2 energy. This is important in our case, since for larger active spaces we use an approximate HCI wave function as the reference, $\refWF$.

To do this, we have primarily considered the same trans-polyacetylene (TPA) system as studied in Sec.~\ref{sec:scaling_len}, for the case with $16$ carbon atoms, $N=16$. We performed the SC-NEVPT2(s) procedure using different HCI wave functions, obtained by varying the HCI threshold, $\epsilon$, which controls the accuracy of the wave function. The exact reference is obtained in the small $\epsilon$ limit.

For TPA (6-31g) results, the following parameters were used. For the norm sampling step of SC-NEVPT2(s), we take $\Nnorm = 950$ and $\Ninit = 50$. For the energy sampling step, $\Nenergy$ is set to $700$, except for $\epsilon = 5 \times 10^{-6}$ Ha where $\Nenergy=1000$. The total residence time is $T = 1.0$, except for $\epsilon = 5 \times 10^{-6}$ Ha where $T=1.5$. We use $20$ burn-in iterations for norm and energy sampling steps.

To address the concern that results may rely on the very small basis set used, we also obtained results for the same TPA system in the cc-pVDZ basis, with two $\epsilon$ values. We also performed a similar analysis for the Fe(II)-Porphyrin (Fe(P)) system. This system and basis is identical to that fully described in Section~\ref{sec:fep}. Results for $\epsilon = 10^{-5}$ Ha are identical to those presented in Section~\ref{sec:fep}. We then performed an additional calculation with $\epsilon = 3 \times 10^{-5}$ Ha.

Results are given in Table~\ref{tab:ref_error}. For each $\epsilon$ value, we state the HCI variational energy, $E^{(0)}$, which is the reference energy in the subsequent SC-NEVPT2 calculation. We also state the Epstein-Nesbet perturbative correction within the CAS (`HCI PT2'), obtained by the semi-stochastic HCI (SHCI) algorithm\cite{Sharma2017}. This gives a measure of error in the reference, but does not include corrections from the FOIS. We then show the SC-NEVPT2(s) energy estimates $E^{(2)}$, and the final energy estimate, obtained as $E^{(0)} + E^{(2)}$.

\begin{table*}[t]
\begin{center}
{\footnotesize
\begin{tabular}{@{\extracolsep{4pt}}lcccccc@{}}
\hline
\hline
 & & & \multicolumn{4}{c}{Energies (Ha)}\\
\cline{4-7}
System & $\epsilon$ (Ha)  & $n_{\mathrm{dets}}$  & HCI variational ($E^{(0)}$)  & HCI PT2  & SC-NEVPT2(s) ($E^{(2)}$)  & Total ($E^{(0)} + E^{(2)}$) \\
\hline
TPA (6-31g)   &  $5 \times 10^{-4}$  &  $1.2 \times 10^{4}$  &  -616.2060  &  -0.0218  & -1.2288(4)  &  -617.4348(4)   \\
              &  $3 \times 10^{-4}$  &  $3.7 \times 10^{4}$  &  -616.2151  &  -0.0169  & -1.2199(5)  &  -617.4351(5)   \\
              &  $1 \times 10^{-4}$  &  $2.4 \times 10^{5}$  &  -616.2341  &  -0.0065  & -1.2035(5)  &  -617.4376(5)   \\
              &  $7 \times 10^{-5}$  &  $3.3 \times 10^{5}$  &  -616.2367  &  -0.0049  & -1.2020(6)  &  -617.4386(6)   \\
              &  $3 \times 10^{-5}$  &  $6.3 \times 10^{5}$  &  -616.2393  &  -0.0034  & -1.1991(5)  &  -617.4384(5)   \\
              &  $5 \times 10^{-6}$  &  $6.4 \times 10^{6}$  &  -616.2439  &  -0.0007  & -1.1938(5)  &  -617.4377(5)   \\
\hline
TPA (cc-pVDZ) &  $1 \times 10^{-4}$  &  $2.2 \times 10^{5}$  &  -616.4946  &  -0.0059  & -1.9154(6)  &  -618.4100(6)   \\
              &  $1 \times 10^{-5}$  &  $2.3 \times 10^{6}$  &  -616.5016  &  -0.0014  & -1.9079(7)  &  -618.4096(7)   \\
\hline
Fe(P)         &  $3 \times 10^{-5}$  &  $2.0 \times 10^{6}$  &  -2245.0225 &  -0.0061  & -3.1708(10) &  -2248.1934(10) \\
              &  $1 \times 10^{-5}$  &  $9.3 \times 10^{6}$  &  -2245.0269 &  -0.0033  & -3.1653(6)  &  -2248.1922(6)  \\
\hline
\hline
\end{tabular}
}
\caption{Results performed for trans-polyacetylene (TPA) with $16$ carbon atoms (C$_{16}$H$_{18}$), and Fe(II)-Porphyrin (Fe(P)) in the $\quintet$ state. We vary the accuracy of the reference wave function, obtained using the HCI method. We then perform SC-NEVPT2(s) using each resulting reference wave function. The final column shows the variation in the total SC-NEVPT2 energy, which is seen to have only weak dependence on the quality of the reference. Even when the reference energy ($E^{(0)}$) is in error by $\sim 38$ mHa, the final SC-NEVPT2 energy is in error by $\sim 3$ mHa, for TPA (6-31g). We also include the HCI PT2 correction within the CAS, obtained by the semi-stochastic HCI approach.}
\label{tab:ref_error}
\end{center}
\end{table*}

The final column can be used to assess the sensitivity of the total SC-NEVPT2 energy to $E^{(0)}$. Interestingly, this total energy shows little variation with $\epsilon$. For TPA (6-31g) with $\epsilon = 5 \times 10^{-4}$ Ha, the HCI variational energy is in error by $\sim 38$ mHa, using only $1.2 \times 10^{4}$ determinants in a $(16\mathrm{e}, 16\mathrm{o})$ active space. However, the final SC-NEVPT2(s) energy is in error by only $\sim 3$ mHa. For $\epsilon = 1 \times 10^{-4}$ Ha, where the reference energy is in error by $\sim 10$ mHa, the total SC-NEVPT2 energy is converged to the exact value within statistical error bars. Similarly, TPA (cc-pVDZ) and Fe(P) results show agreement within error bars after varying $\epsilon$.

These results show that the SHCI PT2 energy (which corrects $E^{(0)}$ itself) should not be included in the final energy estimate. Instead, SC-NEVPT2 energies can be estimated simply as $E^{(0)} + E^{(2)}$. Clearly, including the SHCI PT2 correction would gives energies in a significant error, for the results presented.

The accuracy of $E^{(0)} + E^{(2)}$ can be partially understood, because $E^{(2)}$ is formed as a sum of negative quantities, $\Nlk / (E^{(0)} - \Elk)$. Therefore, as $E^{(0)}$ becomes less negative (larger $\epsilon$), each contribution in the summation becomes more negative. It is not unreasonable to then expect partial cancellation between errors in $E^{(0)}$ and $E^{(2)}$. Nonetheless, the very accurate nature of cancellation here is perhaps surprising. If this result were general, it would be powerful and extremely useful. However, a general statement on the accuracy of this cancellation cannot be made without more testing, for example with several different systems and basis sets, which will be a task for future work. However, these are promising initial results, and justify the HCI wave functions used in the following results sections.

\subsection{Fe(II)-Porphyrin}
\label{sec:fep}

Next we perform calculations of the Fe(II)-Porphyrin (Fe(P)) system. This has been an important benchmark system for multireference methods in recent years, in part due to the difficulty of identifying the spin state ordering\cite{Manni2016, Smith2017, LiManni2018, LiManni2019}. Experimental results on Fe(P) and related systems have usually found the ground state to be a triplet state, although these results are obtained either from a polar solvent or the crystal phase\cite{Kitagawa1979, Mispelter1980, Evangelisti2002, Bartolome2010}. Initial theoretical studies have predicted a quintet $\quintet$ ground state, while a triplet ground state is observed with larger or more careful active space choices\cite{Smith2017, LiManni2018, LiManni2019}. Very recently, it has been suggested that the true ground state is a quintet, when geometrical effects are properly considered\cite{Antalik2020}; we do not consider such effects here.

We focus on a $(32\mathrm{e}, 29\mathrm{o})$ active space used in early studies by Li Manni \emph {et al.}\cite{Manni2016}, and subsequently by Smith \emph{et al}.\cite{Smith2017}. This active space consists of $20$ C $2p_z$, $4$ N $2p_z$ and 5 Fe $3d$ orbitals. We then investigate the effect of dynamic correlation through SC-NEVPT2. In particular, we consider the vertical excitation energy, using the same geometry as Smith \emph{et al.}, which is given in Supplementary Material. This geometry was originally described by Groenhof \emph{et al.}\cite{Groenhof2005}, optimized for the triplet state, and was also used in a DMRG investigation of this system\cite{Olivares2015}. At this fixed geometry, previous results suggest that the ground state is a triplet; for example, this was found to be the case with a larger $(44\mathrm{e}, 44\mathrm{o})$\cite{Smith2017} active space. Lee and co-workers also studied this system recently\cite{Lee2020}, giving a useful summary of recent results, and using auxiliary-field quantum Monte Carlo (AFQMC) to confirm the triplet ground state. However, for this $(32\mathrm{e}, 29\mathrm{o})$ active space, and at the CASSCF level of theory, a $\quintet$ ground state is observed. It is interesting and valuable to investigate to what extent SC-NEVPT2 can correct this situation.

Although Fe(P) has $D_{4h}$ symmetry, we use $D_{2h}$ instead. Using $D_{2h}$ symmetry labels, we calculate the lowest energy states in both the $\quintet$ and $\triplet$ sectors. Note that the irreducible representation $\mathrm{B}_{1\mathrm{g}}$ of $D_{2h}$ corresponds to $\mathrm{A}_{2\mathrm{g}}$ and $\mathrm{B}_{2\mathrm{g}}$ in $D_{4h}$.

The basis set is cc-pVDZ. We use the same CASSCF orbitals optimized by Smith \emph{et al.} for their CASSCF study of this system, where HCI was used as the solver. The reference wave function in SC-NEVPT2(s) was also obtained with HCI, using a final threshold of $\epsilon = 10^{-5}$ Ha, which resulted in a wave function of $\sim 10^7$ determinants for both states. For SC-NEVPT2(s) simulations, parameters used by each process were, for the $\quintet$ state: $\Nnorm = 950$, $\Ninit = 50$, $\Nenergy = 1500$ and $T = 0.4$ (giving $\NElk \approx 72$ on average); and for the $\triplet$ state: $\Nnorm = 900$, $\Ninit = 100$, $\Nenergy = 1260$ and $T = 0.4$ (giving $\NElk \approx 104$ on average); performed with $320$ MPI processes for both states. The following orbitals were frozen in the SC-NEVPT2(s) calculation: $20$ C $1s$, $20$ N $1s$, and $1-3s$, $2-3p$ on the Fe atom, $33$ orbitals in total.

\begin{table}[t]
\begin{center}
{\footnotesize
\begin{tabular}{@{\extracolsep{4pt}}lccc@{}}
\hline
\hline
 & \multicolumn{3}{c}{Energies (Ha)} \\
\cline{2-4}
State & CASSCF & SC-NEVPT2(s) & Total \\
\hline
$\quintet$  & -2245.0269  & -3.1653(6)  &  -2248.1922(6) \\
$\triplet$  & -2244.9957  & -3.1844(7)  &  -2248.1800(7) \\
$\Delta E$  &     0.0312  & -0.0190(9)  &      0.0122(9) \\
\hline
\hline
\end{tabular}
}
\caption{Energies for two low-lying states of Fe(II)-Porphyrin, obtained with CASSCF and SC-NEVPT2(s), using a common geometry for both states. The $(32\mathrm{e}, 29\mathrm{o})$ active space of Li Manni \emph{et al.}\cite{Manni2016} was used. Irreducible representation labels here refer to the $D_{2h}$ point group, which was used for all calculations.}
\label{tab:fe_p}
\end{center}
\end{table}

Results are presented in Table~\ref{tab:fe_p}. Using CASSCF only, the $\quintet$ state is lower in energy than the $\triplet$ state by approximately $31$ mHa. Including the SC-NEVPT2 correction, it is seen that the quintet state remains the ground state, however the energy gap is lowered by approximately $19$ mHa, suggesting an improved result overall.

Note that the CASSCF energy is in error by approximately $+5$ mHa for the $\quintet$ state, and by approximately $+9$ mHa for the $\triplet$ state, due to the finite value of $\epsilon$ used in HCI, although correcting for this does not change our conclusion significantly. It would be simple to improve this by using a smaller value of $\epsilon$, which only has a small effect on the SC-NEVPT2(s) simulation time. This is because coefficients in the reference wave function are obtained by a hash table lookup, the time for which has very weak scaling with the number of determinants.

Our results show that including dynamic correlation through SC-NEVPT2 does noticeably improve the predicted energy gap in this system, but that the expected ordering only occurs with a larger active space. In particular, including the set of $5$ Fe $4d$ orbitals, together with $10$ $\sigma$ bonds between Fe and N atoms ($1$ Fe $4p_x$, $1$ Fe $4p_y$, $4$ N $2p_x$ and $4$ N $2p_y$) results in a $(44\mathrm{e}, 44\mathrm{o})$ active space\cite{Olivares2015, Smith2017}, which gives a triplet ground state. Li Manni \emph{et al.} have also studied a separate model of Fe(P), where C$_{\beta}$H groups are replaced by hydrogen atoms. With this, they also predict a triplet ground state with a more compact $(32\mathrm{e}, 34\mathrm{o})$ active space, which also includes the Fe $4d$ orbitals, and part of the Fe--N $\sigma$ manifold\cite{LiManni2018}. Combined, these results highlight the importance of appropriately choosing the active space in such systems.

\subsection{$ \Cu2O2 $}
\label{sec:cu2o2}

\begin{table}[t]
\begin{center}
{\footnotesize
\begin{tabular}{@{\extracolsep{4pt}}lcccccc@{}}
\hline
\hline
 & \multicolumn{6}{c}{F}\\
\cline{2-7}
Method  & $0$  & $0.2$  & $0.4$  & $0.6$  & $0.8$  & $1$ \\
\hline
HCI-SCF                  &  22.4    & 14.3    & 8.2     & 3.7     & 1.0    & 0 \\
SC-NEVPT2(s)             &  41.3(8) & 33.5(9) & 26.3(9) & 19.9(8) & 9.3(9) & 0 \\
CAS(16,14)${}^a$         &  0.2     & -7.2    & -12.7   & -16.3   & -14.0  & 0 \\
CR-CCSD(TQ)${}^a$        &  35.1    & 26.7    & 18.9    & 10.7    & 3.1    & 0 \\
CR-CCSD(TQ)${}_L$${}^a$  &  38.5    & 28.8    & 20.0    & 11.4    & 3.6    & 0 \\
DMRG-CI${}^b$            &  -12.8   & -20.9   & -21.5   & -16.7   & -10.0  & 0 \\
DMRG-SCF${}^b$           &  26.4    & 17.9    & 11.0    & 5.1     & 1.1    & 0 \\
DMRG-SC-CTSD${}^b$       &  37.4    & 29.0    & 22.0    & 14.4    & 6.1    & 0 \\
\hline
\hline
\end{tabular}
}
\caption{Energies (in kcal mol${}^{-1}$) from various methods, including SC-NEVPT2(s), for the isomerization of $ \Cu2O2 $ between bis($\mu$-oxo) and $\mu$-$\eta^2$:$\eta^2$-peroxo isomers. Energies are relative to the $\mu$-$\eta^2$:$\eta^2$-peroxo isomer ($F = 1.0$). Results labelled $a$ are from Ref.~\onlinecite{Cramer2006}. Results labelled $b$ are from Ref.~\onlinecite{Yanai2010}. Note that we use a different basis set to these two studies.}
\label{tab:cu2o2}
\end{center}
\end{table}

\begin{figure}[t]
\includegraphics[width=\linewidth]{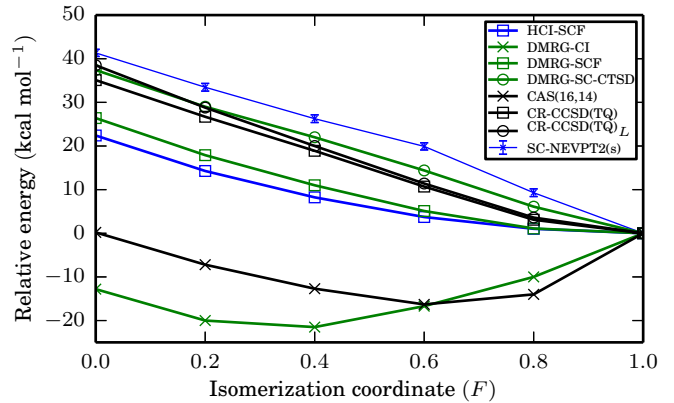}
\caption{Energies for the isomerization of $ \Cu2O2 $, relative to the $\mu$-$\eta^2$:$\eta^2$-peroxo isomer ($F = 1.0$). Data plotted is the same as in Table~\ref{tab:cu2o2}. Results plotted in black are from Ref.~\onlinecite{Cramer2006}. Results plotted in green are from Ref.~\onlinecite{Yanai2010}.}
\label{fig:cu2o2}
\end{figure}

As a final example, we consider the $ \Cu2O2 $ molecule. In particular, we study the isomerization between bis($\mu$-oxo) and $\mu$-$\eta^2$:$\eta^2$-peroxo isomers. This model and process, in particular when combined with appropriate ligands, has an important role as an active site for O$_2$ activation by enzymes such as tyrosinase. Given the presence of transition metals, it is expected that treatment of static correlation may be important, and it has further been suggested that a balanced treatment of static and dynamic is required for accurate results. Moreover, existing benchmarks are available from previous computational studies\cite{Cramer2006, Yanai2010}, making this a sensible test system.

We describe the isomerization process using the same geometries of Cramer \emph{et al.}\cite{Cramer2006} In this, the Cu--Cu distance is equal to $2.8 + 0.8F$ \AA, while the O--O distance is equal to $2.3 - 0.9F$ \AA. Here, $F$ is a parameter which varies from $0$ to $1$. $F=0$ indicates the bis($\mu$-oxo) geometry and $F=1$ indicates the $\mu$-$\eta^2$:$\eta^2$-peroxo geometry.

We use the ANO-RCC-VQZP basis set\cite{Roos2004, Roos2005}, which corresponds to Cu:[$21s15p10d6f4g2h/7s6p4d3f2g1h$] and O:[$14s9p4d3f2g/4s3p2d2f1g$] contractions. This is slightly different to the basis used in other studies, such as that by Yanai \emph{et al.}\cite{Yanai2010}

We take the same $(28\mathrm{e}, 32\mathrm{o})$ active space of Yanai \emph{et al.}, consisting of all Cu $3d$ and $4d$ orbitals, and all O $2p$ and $3p$ orbitals.

CASSCF orbitals are obtained with HCI using a final threshold of $\epsilon = 10^{-4}$ Ha. We then use a tighter threshold of $\epsilon = 2 \times 10^{-5}$ Ha to generate the reference wave function for SC-NEVPT2(s). This results in HCI wave functions with between $2.0\ \times 10^{7}$ and $2.6 \times 10^{7}$ determinants, depending on $F$. We then perform SC-NEVPT2(s), using norm parameters $\Nnorm = 450$, $\Ninit = 50$. The number of energy samples, $\Nenergy$, is between $2000$ and $2100$, and simulations were run with either $320$ or $360$ processes, depending on the value of $F$. The total residence time $T$ was set to $0.4$, which gave $\NElk$ between $50$ and $55$ on average (in addition to $50$ burn-in iterations).

Results are given in Table~\ref{tab:cu2o2}, and plotted in Fig.~\ref{fig:cu2o2}. We also include results from previous studies for comparison. In particular, CAS(16,14), CR-CCSD(TQ) and CR-CCSD(TQ)${}_L$ results were taken from the study of Cramer \emph{et al.}\cite{Cramer2006} and DMRG-CI, DMRG-SCF and DMRG-SC-CTSD results were taken from the study of Yanai \emph{et al.}\cite{Yanai2010} Our CASSCF results, obtained using HCI as a solver, are labelled `HCI-SCF'. It is known to be difficult to obtain the correct isomerization profile for this system. Too small an active space leads to an unphysical minimum. HCI-SCF results show that the more substantial $(28\mathrm{e}, 32\mathrm{o})$ active space removes this minimum, as previously found by Yanai using DMRG-SCF. More accurate results are obtained when dynamical correlation is included. Our SC-NEVPT2(s) results are approximately in agreement with existing results. We find slightly larger relative energies than previous results. However, we use a larger basis set, so it is perhaps expected that results will not be identical. Cramer \emph{et al.} also use a pseudopotential for Cu atoms, while we freeze core electrons. Overall, these results show reasonable agreement, and demonstrate the usefulness of this approach for a significant active space.

\section{Conclusion}
\label{sec:conclusion}

In this work we have developed a stochastic approach to performing strongly contracted NEVPT2. This method reproduces exact SC-NEVPT2 energies within statistical error bars, but avoids the prohibitive cost of constructing and storing $3$ and $4$-body RDMs.

The method has low scaling with the number of virtual orbitals, $n_v$. The cost to sample a fixed number of perturber energies, $\Elk$, is independent of $n_v$, while the increase in associated statistical error is low for small basis sets, plateauing off for larger basis sets. The scaling with number of active space orbitals is more restrictive. In particular, we investigated the scaling of the overall computational cost with molecular size, $N$, for polyacetylene molecules. In this case, the number of core, active and virtual orbitals all increase linearly with $N$, and the total cost (after accounting for increase in statistical error) was found to scale roughly as $\mathcal{O}(N^{8.2})$.

We also investigated the sensitivity of the final SC-NEVPT2 energy to a reference wave function of varying accuracy. Interestingly, we found final energies to remain accurate, with relatively weak dependence on the quality of the reference energy. If this result were general then it would be very powerful. We intend to study this for further systems to investigate this possibility.

The method was applied to example systems where multi-reference behaviour is expected to be important: Fe(II)-Porphyrin with a $(32\mathrm{e}, 29\mathrm{o})$ active space, and $ \Cu2O2 $ with a $(28\mathrm{e}, 32\mathrm{o})$ active space. The method was successfully applied to these large active spaces, raising the possibility of obtaining SC-NEVPT2 results, without approximations, in larger active spaces than previously considered. These calculations were performed with moderate computer resources. However, the approach has good parallel efficiency, such that it could be used in a straightforward manner on much larger parallel computers, as have been used in many QMC studies previously.

There are several areas in which this method could be developed. First, it will be important to develop SC-NEVPT2(s) to work with other types of wave functions, in particular VMC wave functions. Such wave functions can be well suited to strong correlation, and with favorable scaling\cite{Casula2004, Neuscamman2012, Neuscamman2013}. Because only wave function overlaps ($\bra n | \refWFn \ket$) and 1- and 2-RDMs are needed, this should be a straightforward task. Our code already supports optimization of VMC wave functions, including calculation of the required overlaps and RDMs. Second, we are keen to investigate approaches to reduce the scaling, in particular with respect to active space size. With these developments, we hope that this may be a robust method to perform NEVPT2 with active spaces of $40$ to $50$ orbitals, which we believe would be valuable in the general task of performing strongly correlated electronic structure calculations.

\section*{Supplementary Material}
Supplementary material includes the geometry of the Fe(II)-Porphyrin model studied in this article. This geometry was taken from Ref.~(\onlinecite{Groenhof2005}). The geometries for all other systems are stated in the article.

\begin{acknowledgments}
NSB is grateful to St John's College, Cambridge for funding and supporting this work through a Research Fellowship. SS and AM were supported by NSF through the grant CHE-1800584. SS was also partly supported through the Sloan research fellowship. This study made use of the CSD3 Peta4-Skylake CPU cluster at the University of Cambridge, and the Summit supercomputer at CU Boulder.
\end{acknowledgments}

\section*{Data availability statement}

The data that supports the findings of this study are available within the article.

\appendix

\section{Potential biases}
\label{sec:biases}

Because a large number of energies $\Elk$ must be sampled, each with its own independent random walk, only a limited number of samples can be used to estimate each $\Elk$. This is different to the typical case in VMC, where a single energy is to be estimated by a long random walk (typically by the Metropolis algorithm). In general, systematic biases in QMC will become larger as the number of samples becomes smaller. Therefore, there are some potential biases to consider carefully for the algorithm presented.

\subsection{Burn-in}

Each random walk with the CTMC algorithm has a burn-in period. In practice, we have found that results are essentially identical regardless of whether burn-in iterations are discarded or not, suggesting this to be a negligible effect here. Nonetheless, it is sensible to account for this possibility where affordable. We therefore typically discard the first $50$ iterations for each random walk, both in $S_0^{(0)}$ (for $\Nlk$ estimation) and in each $\Slk$ sampled (for $\Elk$ estimation).

\subsection{CTMC estimates of $\Elk$}

Some care is required in using the CTMC algorithm. In CTMC, a sample from a given determinant $|n\ket$ is weighted by a corresponding residence time, defined as $t_n = \frac{1}{\sum_p r(p \leftarrow n)}$. The final point estimate of $\Elk$ is obtained by
\begin{equation}
\hat{E}_l^{(k)} = \frac{ \sum_n t_n \ELDn }{ \sum_n t_n }.
\label{eq:weighted}
\end{equation}
Using a constant number of iterations for each $\Elk$ leads to small systematic error, which becomes noticeable for very large systems. Instead, each $\Elk$ should be estimated with a constant total residence time, $T = \sum_n t_n$. We therefore run CTMC random walks until some fixed threshold time is reached, at which point the walk is ended. This is found to resolve all such issues with systematic errors in $\Elk$ estimates.

\subsection{Bias in $(\refE-\Elk)^{-1}$ estimator}

Contributions to $E_m^{(2)}$ each take the form $(E_m^{(0)} - \Elk)^{-1}$, where each $\Elk$ is stochastically sampled. Even if the estimator for $\Elk$ is unbiased, the final result will be biased because $\mathrm{E}[\frac{1}{X}] \ne \frac{1}{\mathrm{E}[X]}$. Estimators of this type are very common in QMC, and associated biases are typically negligible. In the current case, however, the bias is larger because the number of samples used to estimate each $\Elk$ is very small ($\sim 50 - 100$), for the calculations presented in this work.

To see the issue more clearly, we can consider a Taylor expansion of $(E_m^{(0)} - \ElkHat)^{-1}$, where $\ElkHat$ is a point estimate of $\Elk$. We may write $\ElkHat = \Elk + \delta$, where $\delta$ denotes the error. Assuming that $\ElkHat$ is unbiased, we have that $\mathrm{E}[\delta] = 0$. One can then write
\begin{align}
\frac{1}{E_m^{(0)} - \ElkHat} &= \frac{1}{E_m^{(0)} - \Elk - \delta}, \\
            &= \frac{1}{(E_m^{(0)} - \Elk) \Big[1 - \frac{\delta}{E_m^{(0)} - \Elk} \Big] }, \\
            &= \frac{1}{E_m^{(0)} - \Elk} \Big[ 1 + \frac{\delta}{E_m^{(0)} - \Elk} \nonumber \\
            &\qquad\qquad+ \frac{\delta^2}{(E_m^{(0)} - \Elk)^2} + \mathcal{O}(\delta^3) \Big].
\end{align}
We can use use this to look at the expected value of $(\refE - \ElkHat)^{-1}$:
\begin{align}
\mathrm{E} \bigg[ \frac{1}{\refE - \ElkHat} \bigg] &= \frac{1}{E_m^{(0)} - \Elk} \Big[ 1 + \frac{\mathrm{E}[\delta^2]}{(E_m^{(0)} - \Elk)^2} \nonumber \\
      &\qquad\qquad\qquad\quad\enspace+ \mathcal{O}(\delta^3) \Big], \\
  &= \frac{1}{E_m^{(0)} - \Elk} + \frac{\mathrm{var}[\ElkHat]}{(E_m^{(0)} - \Elk)^3} \nonumber \\
      &\qquad\qquad\qquad\quad+ \mathcal{O}(\delta^3).
\end{align}
Therefore, it can be seen that the bias will increase as the energy difference $E_m^{(0)} - \Elk$ becomes smaller, and as the estimate of $\Elk$ becomes more noisy.

The above gives an expression to correct much of the bias:
\begin{equation}
E_{\mathrm{bias \: corr.}} = - \frac{\mathrm{var}[\ElkHat]}{(E_m^{(0)} - \Elk)^3}.
\label{eq:bias_corr}
\end{equation}
Using this expression requires an estimate of the variance of $\ElkHat$. If the Metropolis algorithm were used, the standard estimator for the variance of the mean would be used:
\begin{equation}
\hat{\sigma}_{\hat{E}_l^{(k)}}^2 = \frac{1}{N_s(N_s - 1)} \sum_n ( \ELDn - \ELDmean )^2,
\end{equation}
where $N_s$ is the number of samples, and $\ELDmean$ the sample mean. Instead, we use the CTMC algorithm, where the estimator for $\Elk$ is formed as a weighted sum, as in Eq.~(\ref{eq:weighted}). An estimator for the variance of a weighted sum is more complicated, and there is no generally accepted formula for all applications. We have tested several estimators, and found that the following formula\cite{Cochran1977, Gatz1995} is very accurate for our case, which we therefore use:
\begin{align}
\hat{\sigma}_{\hat{E}_l^{(k)}}^2 &= \frac{1}{T} \frac{N_s}{(N_s - 1)} \Big[ \sum_n ( t_n \ELDn - T \ELDmean )^2 \\
                           &- 2 \ELDmean \sum_n (t_n - T)(t_n \ELDn - T \ELDmean ) \\
                           &+ \ELDmean^2 \sum_n ( t_n - T )^2 \Big].
\label{eq:variance}
\end{align}
Here, $T = \sum_n t_n$ is the total residence time, and $t_n$ act as weights in the estimator for $E_l^k$, as in Eq.~(\ref{eq:weighted}). $\ELDn$ is the local energy with respect to the Dyall Hamiltonian, as in Eq.~(\ref{eq:local_dyall}). In addition, samples $\ELDn$ are serially correlated, and we account for this by using an automated reblocking procedure\cite{Flyvbjerg1989}.

\subsection{Example: N$_2$ cc-pVDZ}

As a simple example to demonstrate these concepts, in particular the estimation of $\sigma_{\hat{E}_l^{(k)}}^2$ and the bias correction term, we consider N$_2$ in a cc-pVDZ basis set. This is the same example considered in Sec.~\ref{sec:scaling_vir}, using a $(10\mathrm{e}, 8\mathrm{o})$ active space and 2 core orbitals.

We consider the estimation of a single perturber energy, $\Elk$, of type $vv$, involving the two virtual orbitals that are lowest in energy. For this small example, it is possible to enumerate all determinants in $\Slk$ and calculate the exact $\Elk$. By repeating the stochastic estimation of $\Elk$ a large number of times, we can investigate the above effects. In particular, we repeat this estimation of $\Elk$ $100,000$ times, so that we can accurately construct the distribution function and investigate the true variance and bias.

For the perturber in question, the exact result is $\Elk = -105.25896$ Ha. Performing the CTMC estimation of $\Elk$, exactly as in the SC-NEVPT2(s) algorithm, and then averaging over the $100,000$ repeated estimates, gives $\Elk = -105.25882(15)$ Ha, so that the method is unbiased within error bars. An accurate estimate of the variance (obtained directly from the constructed probability distribution) is $\mathrm{Var}[\hat{E}_l^{(k)}] = 0.00229$ Ha${}^2$, while the estimate from Eq.~(\ref{eq:variance}) is $\sigma_{\hat{E}_l^{(k)}}^2 = 0.00233$ Ha${}^2$.

Similarly, the difference between the exact and estimated values of $(\refE - \Elk)^{-1}$ is $3.3(11) \times 10^{-5}$ Ha$^{-1}$, indicating the possibility of a small bias. Including the above bias correction changes this discrepancy to $-1.1(11) \times 10^{-5}$ Ha$^{-1}$, suggesting an improvement. In this case the correction is extremely small, so could be ignored. For non-trivial problems this correction needs more careful consideration. For the $ \Cu2O2 $ examples in Sec.~\ref{sec:cu2o2}, the bias correction in Eq.~(\ref{eq:bias_corr}) is of size $\approx 0.6$ mHa, for each value of $F$. We therefore include this correction term in all results presented in this article.

%

\end{document}